# Transverse spinning of light with globally unique handedness

Xianji Piao, Sunkyu Yu, and Namkyoo Park*

*Photonic Systems Laboratory, Department of Electrical and Computer Engineering, Seoul National University, Seoul 08826, Korea*

*E-mail address for correspondence: nkpark@snu.ac.kr

**Access to the transverse spin of light[1,2] has unlocked new regimes in topological photonics[3,4] and optomechanics[5-7]. To achieve the transverse spin of nonzero longitudinal fields, various platforms that derive transversely confined waves based on focusing[7-9], interference[10], or evanescent waves[3,11,12] have been suggested. Nonetheless, because of the transverse confinement inherently accompanying sign reversal of the field derivative, the resulting transverse spin handedness experiences spatial inversion[1,2], which leads to a mismatch between the densities of the wavefunction and its spin component and hinders the global observation of the transverse spin. Here, we reveal a globally pure transverse spin in which the wavefunction density signifies the spin distribution, by employing inverse molding of the eigenmode in the spin basis. Starting from the target spin profile, we analytically obtain the potential landscape and then show that the elliptic-hyperbolic transition around the epsilon-near-zero permittivity allows for the global conservation of transverse spin handedness across the topological interface between anisotropic metamaterials. Extending to the non-Hermitian regime, we also develop annihilated transverse spin modes to cover the entire Poincaré sphere of the meridional plane. Our results enable the complete transfer of optical energy to transverse spinning motions and realize the classical analogy of 3-dimensional quantum spin states.**

The very nature of transverse waves of light has hindered our access to the rich physics and applications related to longitudinal fields. However, the recent emergence of plasmonics[13] and metamaterials[14], which provide superior light confinement, has lifted the restriction on longitudinal fields, and unconventional dynamics, including left-handed metamaterials[15], artificial magnetics[16], full-field reconstruction[17], and plasmonic radiation[18], can now be realized. By controlling the phase of the longitudinal field relative to the transverse component, the hidden optical quantity of transverse spin angular momentum[1,2] that overcomes the transverse nature of light has also been realized. Intensive studies were subsequently conducted to examine and exploit the optical transverse spin and offer unexplored degrees of freedom for 3-dimensional light-matter interactions[3-6] for applications in optomechanics[5-7] and topological photonics[3,4].

Although the platforms for optical transverse spin (T-spin) can be classified according to the excitation method of the longitudinal wave, which is either a propagating[7-9] or an evanescent wave[3,11,12], the necessary condition for both cases is identical: the transverse confinement of the light wave that drives the nonzero longitudinal component of the field. Nonetheless, compared with the longitudinal spin, which can be defined as a globally pure and exclusive quantity that constitutes the circular polarization of light, the T-spin usually remains a local quantity[1,2,6-11], which leads to a separation between the densities of the wavefunction $|\Psi|^2 = |\psi_+ \cdot \mathbf{e}_+ + \psi_- \cdot \mathbf{e}_-|^2$ and each of its spin components, $|\psi_+|^2$ or $|\psi_-|^2$. Considering the current absence of T-spin filters that are analogous to circular polarizers for the longitudinal spin, the realization of a globally pure T-spin mode that achieves the relation of $|\Psi|^2 = |\psi_+|^2$ is strongly desired. Such a T-spin mode will allow for the utilization of the wavefunction density $|\Psi|^2$ as the direct observable of the transverse spin $\psi_+$, in the context of far-field transport and T-spin mode observations.

In this paper, we propose a realization of pure T-spin modes that preserve unique handedness in a global domain. Noticing that the spatial inversion of the transverse spin originates from the transverse confinement of a field, we first reveal that the globally unique handedness of T-spin can be achieved by implementing a topological transition around the uniaxial epsilon-near-zero (ENZ) permittivity in the transverse coordinate. Then, by molding the potential landscape according to the target spin wave profile, we demonstrate the construction of arbitrary T-spin modes, including zero spin, which preserves the global spin purity. Our results achieve almost complete coverage of the Poincaré sphere including the longitudinal component, and allow the designer manipulation of the transverse spin mode and its far-field transport, thereby paving the way toward full access to the 3-dimensional polarization states of propagating light.

Without a loss of generality, we focus on the transverse magnetic (TM) mode for the analysis of T-spin modes, and it can be readily extended to the transverse electric (TE) mode with dual symmetry[19]. Conventional T-spin realizations in isotropic media (Fig. 1a,b) are then derived from the confinement of light by exploiting waves focused in bulk media (Fig. 1a) or evanescent waves at the interface (Fig. 1b). Although considerable attention has been focused on methods of light *confinement* for the T-spin excitation, in this study, we investigate the role of the 'longitudinal' displacement field that originates from the confinement. We stress that the longitudinal component of the displacement vector ($D_y = i \cdot \partial_x H_z / \omega$) always accompanies the spatial sign inversion along the transverse axis $x$ to achieve the confinement ($\partial_x H_z(x < 0) > 0$ and $\partial_x H_z(x > 0) < 0$) as shown in Fig. 1. Because $D_x$ is continuous, this sign inversion of $D_y$ leads to the inevitable sign inversion of the T-spin density along the transverse axis regardless of prior realization methods that utilize isotropic media (Fig. 1a,b) and results in $|\Psi|^2 \neq |\psi_+|^2$ (Fig. 1d,e).

The aforementioned sign reversals of the **D**-field and the resulting spin handedness of the **E**-field can be overcome via the constitutive relation $\mathbf{E} = \boldsymbol{\varepsilon}^{-1}\mathbf{D}$. To release the restriction on the **D**-field, we consider anisotropic materials of $\boldsymbol{\varepsilon}(x)$, which adds new degrees of freedom to the control of the electric field irrespective of the displacement field. For example, with $\varepsilon_y (x < 0) > 0$ and $\varepsilon_y (x > 0) < 0$, the globally unique handedness of the spin $|\Psi|^2 = |\psi_+|^2$ (Fig. 1c) could be achieved, where its profile could also be further controlled by $\varepsilon_x(x)$ (Fig. 1f-i).

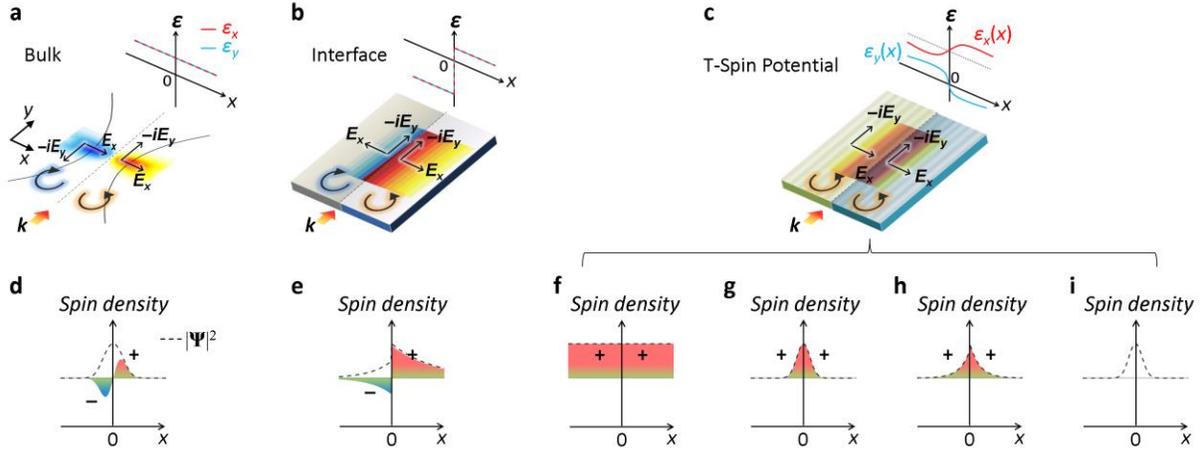

**Figure 1. Illustration of a globally pure T-spin mode.** (**a,b**) Conventional realizations of T-spin from (**a**) a Gaussian focused wave propagating in a homogeneous isotropic bulk medium ($\varepsilon_x = \varepsilon_y > 0$ for all $x$), and (**b**) an evanescent plasmonic wave confined at an interface between an isotropic metal ($\varepsilon_x = \varepsilon_y < 0$ for $x < 0$) and a dielectric medium ($\varepsilon_x = \varepsilon_y > 0$ for $x > 0$). (**c**) Proposed T-spin potential between two inhomogeneous anisotropic media around the uniaxial ENZ medium. The overlapping fields in (**a-c**) indicate each spin density distribution. (**d-i**) Corresponding spin density profiles $|\psi_+(x)|^2 - |\psi_-(x)|^2$ of (**a-c**) compared with the entire wavefunction density $|\Psi(x)|^2$ (dashed lines): (**d**) Gaussian wave, (**e**) plasmonic wave, (**f-h**) T-spin waves with globally pure handedness ($|\Psi(x)|^2 = |\psi_+|^2$) in the forms of (**f**) a plane wave, (**g**) a Gaussian wave, and (**h**) an evanescent wave, and (**i**) an annihilated T-spin wave without handedness ($|\psi_+(x)|^2 = |\psi_-(x)|^2$). The spatial inversion of spin handedness occurs in (**d,e**).

In the context of the inverse design of optical potentials from the target eigenmode, the T-spin mode of globally 'pure' handedness in the entire space ($\psi_- = 0$ for all regions) can be molded. We assume electrically anisotropic materials having 1-dimensional (1D) variations of $\varepsilon_{x,y}(x)$ and a nonmagnetic feature ($\mu = \mu_0$). The electric field then has the form of $\mathbf{E}(x,y) = \mathbf{\Psi}(x) \cdot e^{-i\beta y}$, where $\mathbf{\Psi}(x)$ is the wavefunction envelope and $\beta$ is the $y$-propagating wavevector. In the T-spin representation of $\mathbf{\Psi}(x) = \psi_+(x) \cdot \mathbf{e}_+ + \psi_-(x) \cdot \mathbf{e}_-$ for the $z$-axis spins $\mathbf{e}_\pm = (\mathbf{e}_x \pm i \cdot \mathbf{e}_y)/2^{1/2}$ 'transverse' to $y$-propagating waves, a globally positive T-spin mode $\mathbf{\Psi}(x) = \psi_+(x) \cdot \mathbf{e}_+$ from $\psi_-(x) = 0$ is then achieved with the anisotropic and inhomogeneous optical potential (see Supplementary Note S1 for the detailed derivation)

$$\varepsilon_x(x) = \varepsilon_e + \frac{\sqrt{\varepsilon_e}}{k_0} \cdot \partial_x \left( \log \psi_+ \right)$$
$$\varepsilon_y(x) = -\frac{1}{k_0^2} \cdot \frac{\partial_x^2 \psi_+}{\psi_+} - \frac{\sqrt{\varepsilon_e}}{k_0} \cdot \partial_x \left( \log \psi_+ \right)$$
(1)

where $k_0 = 2\pi/\lambda_0$ is the free space wavenumber and $\varepsilon_e$ is the effective permittivity of the +$y$-propagating ($\beta > 0$) T-spin eigenmode from $\beta^2 = \varepsilon_e \cdot k_0^2$ (see Supplementary Note S2 for the dependence on the propagation direction and mirror symmetry). Equation (1) can be applied to obtain potential landscapes for any nodeless spatial profiles of the target $\psi_+(x)$.

First, we consider the trivial case of a T-spin 'plane wave' where $\partial_x \psi_+ = 0$. The required potential profile from Eq. (1) then becomes constant as $\varepsilon_x(x) = \varepsilon_e$ and $\varepsilon_y(x) = 0$. This solution reveals that zero longitudinal permittivity $\varepsilon_y = 0$, which corresponds to uniaxial ENZ[20-22] materials, imposes the full degree of freedom on the longitudinal electric field $E_y$ and thus allows the emergence of the ideal T-spin plane wave ($D_y = \varepsilon_y \cdot E_y = 0$). However, from $\varepsilon_y(x) = 0$, any $E_y(x)$ is able to satisfy Maxwell's equations, thereby hindering the exclusive excitation of the pure T-spin state. We thus explore the condition of nontrivial "confined" eigenmodes that have T-spin polarizations distinct from that of the trivial plane wave case.

Without a loss of generality, we consider an example of the Gaussian profile $\psi_+(x) = exp(-x^2/(2\sigma^2))$ while keeping $\psi_-(x) = 0$ (Fig. 2a, $\sigma = 5\lambda_0$). The required optical potential solution (Fig. 2b) is then 'inhomogeneous' anisotropic media (see Supplementary Note S3 for the field profiles and power flow of the target eigenmode $\Psi(x) = \psi_+(x) \cdot \mathbf{e}_+$ with different modal sizes). Remarkably, although $\varepsilon_x(x)$ has its distribution near the target effective permittivity $\varepsilon_e$ (upper black dashed line in Fig. 2b), $\varepsilon_y$ exhibits an intriguing transition from negative to positive values near $x = 0$, thus realizing the topological transition[23] of the isofrequency contour (IFC) between hyperbolic[24] and elliptic materials (Fig. 2c). Compared with the plane wave solution of constants $\varepsilon_x(x) = \varepsilon_e$ and $\varepsilon_y(x) = 0$, the spatial transition of the IFC near uniaxial ENZ ($\varepsilon_y(x < 0) > 0$ and $\varepsilon_y(x > 0) < 0$) not only allows the pure T-spin handedness of the target eigenmode (Fig. 2d) but also yields a momentum mismatch (yellow regions in Fig. 2c), which leads to the confinement of the eigenmode near $x \sim 0$ (Fig. 2e). We also note that the target T-spin eigenmode is separated from the other eigenmodes in both the spatial and momentum domains (see Supplementary Note S4 for details on the nearby eigenmodes), which enables the exclusive excitation of the T-spin without the other polarization states.

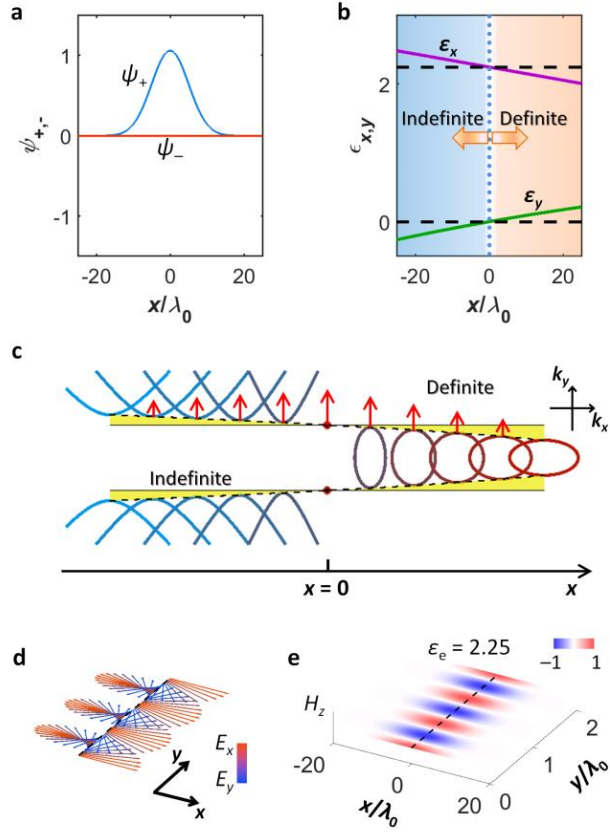

**Figure 2. Example of the pure T-spin mode in anisotropic and inhomogeneous media.** (**a**) Target spin profiles of the eigenmode that exhibit the Gaussian $\psi_+(x)$ and the suppressed $\psi_-(x)$. (**b**) Inverse design-obtained anisotropic permittivities $\varepsilon_{x,y}(x)$ for the target eigenmode. The background colors indicate the topology of the materials (blue for hyperbolic media and red for elliptic media). Black dotted lines each indicate the target effective permittivity $\varepsilon_e = 2.25$ and the zero permittivity. (**c**) Schematic of the spatial variation of the IFC for the designed potential in (**b**). Red arrows show the direction of the group velocity at each position. The yellow region depicts the spatially varying momentum mismatch. (**d**) Electric field evolution $E$ at $x = 0$ and (**e**) normalized magnetic field ($H_z$) distribution. The color bars in (**d**) represent the ratio of transverse ($E_x$) and longitudinal ($E_y$) electric fields. The results of (**d,e**) are obtained using the eigenmodal analysis of COMSOL Multiphysics.

Although the result in Fig. 2 demonstrates the existence of a confined T-spin mode with a globally

'pure' handedness, the spatially varying condition of anisotropic materials may hinder practical implementations. Because the modal profile of $e^{-\alpha x}$ allows for 'homogeneous' realizations (Eq. (1) of $\varepsilon_x(x) = \varepsilon_e$ and $\varepsilon_y(x) = -\alpha^2/k_0^2$), we now explore the condition of potential landscapes for the confined modal profiles of $\psi_+(x) = exp(-|x|^g/2\sigma^g)$ (Fig. 3a,b, $g = 2$ to 1, from Gaussian to exponential). As expected, the potential becomes homogeneous for $g \sim 1$ except for the discontinuity in $\partial_x \psi_+$ at $x = 0$ and the corresponding singularity of $\varepsilon_y(x)$. By neglecting the contribution of the second-order derivative $\partial_x^2 \psi_+$ in Eq. (1) for the ansatz function $\psi_+(x) \sim e^{-\alpha|x|}$, Eq. (1) can be used to derive the topological-transition interface between hyperbolic $[\varepsilon_x, \varepsilon_y] = [\varepsilon_e + \Delta\varepsilon, -\Delta\varepsilon]$ and elliptic $[\varepsilon_x, \varepsilon_y] = [\varepsilon_e - \Delta\varepsilon, \Delta\varepsilon]$ materials ($\Delta\varepsilon = (\varepsilon_e^{1/2}/k_0) \cdot \alpha > 0$, Fig. 3c; also, see the Methods Summary for the derivation of the interface mode), where the value of $\Delta\varepsilon$ determines the confinement of the eigenmode $\psi_+(x) \sim e^{-\alpha|x|}$. Figure 3d represents the electric field distribution of the T-spin interface mode for the material parameters $\varepsilon_e = 2$ and $\Delta\varepsilon = 0.5$. As expected, the signs of the transverse electric field $E_x$ and the following spin density are preserved along the entire $x$-axis, achieving $|\psi_+| \gg |\psi_-|$ (Fig. 3e). Moreover, the $\Delta\varepsilon$-dependent discontinuity in the transverse field $E_x$ (red lines) indicates that a negligible excitation of $\psi_-$ occurs except in the case of deep-subwavelength confinements (see Supplementary Note S5). Figures 3f and 3g show the global excitation of the T-spin mode $\psi_+$ and suppression of $\psi_-$ at the interface that employs the platform of hyperbolic[24] and elliptic layered metamaterials, with an oblique plane wave incidence (see Supplementary Note S6 for the detailed structural parameters and Supplementary Note S7 for excitation conditions). Along with the metamaterial platform, the homogeneous realization also opens up the possibility of T-spin mode excitation at the interface in between natural anisotropic media[25,26].

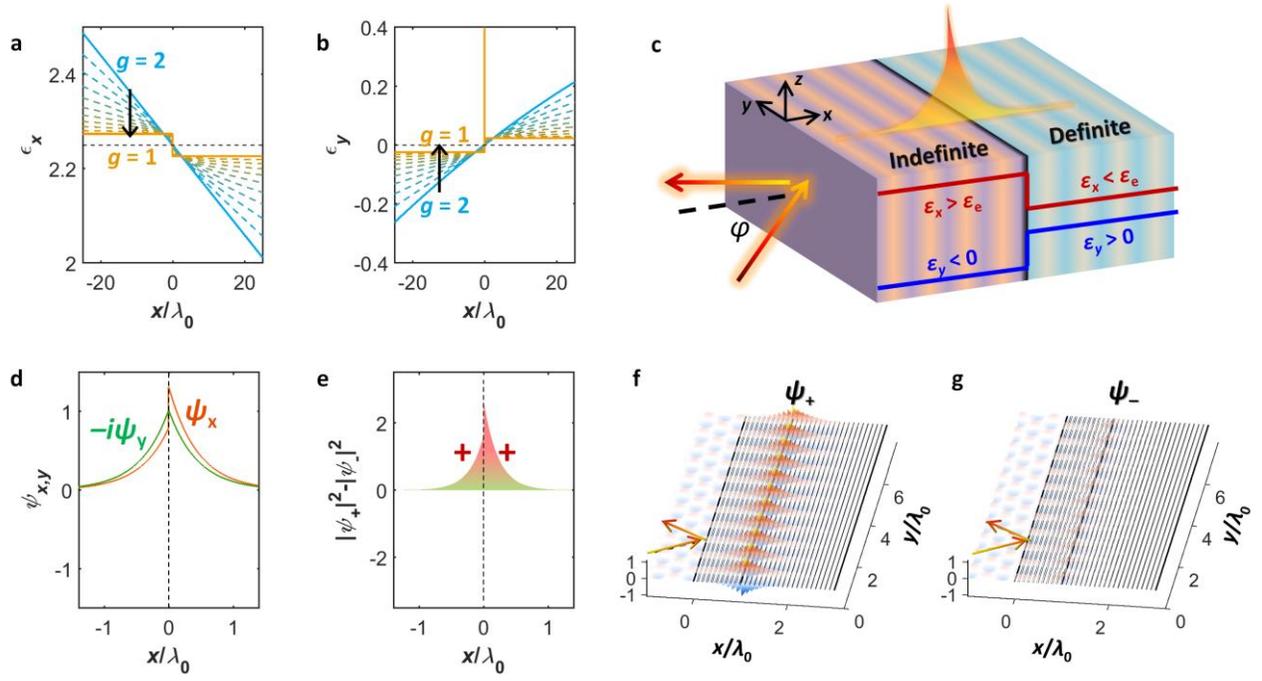

**Figure 3. Pure T-spin interface mode using homogeneous materials of hyperbolic and elliptic IFC.** Material parameters (**a**) $\varepsilon_x(x)$ and (**b**) $\varepsilon_y(x)$ required for the target T-spin eigenmode $\psi_+(x) = exp(-|x|^g/2\sigma^g)$ ($g = 2$ to 1, blue to orange). (**c**) Schematic for the excitation of the T-spin interface mode based on homogeneous media. The colored arrows represent the incident and reflected waves used for the excitation of the *y*-propagating T-spin interface mode. Red and blue lines indicate the arrangement of anisotropic permittivities. $\varphi$ is the incident angle. (**d**) Envelope functions of the electric field $\psi_{x,y}$ ($E_{x,y}(x,y) = \psi_{x,y}(x)\cdot e^{-i\beta y}$) and (**e**) net spin density $|\psi_+|^2 - |\psi_-|^2$ of the T-spin interface mode. (**f,g**) Excitation of the T-spin interface mode implemented in the platform of hyperbolic- and elliptic- layered metamaterials through the oblique incidence ($\varphi = 43.2°$): (**f**) excited $\psi_+$ and (**g**) $\psi_-$ fields. For all cases, the transverse permittivity is $\varepsilon_x(x < 0) = 2.5$ and $\varepsilon_x(x \geq 0) = 1.5$, and the longitudinal permittivity is $\varepsilon_y(x < 0) = -0.5$ and $\varepsilon_y(x \geq 0) = 0.5$, corresponding to $\varepsilon_e = 2$ and $\Delta\varepsilon = 0.5$. The detailed design of the layered metamaterials and excitation conditions in (**f,g**) are shown in Supplementary Notes S6,7. The results of (**f,g**) are obtained from the stable scattering matrix calculation[27].

Extending the realization of globally pure T-spin waves further, we challenge the feasibility of obtaining a fundamental but unexplored state of polarization (SOP): the zero-T-spin mode corresponding to a linear polarization 'oblique' to the propagation direction on the Poincaré sphere of the meridional plane (Fig. 4a), which we call the T-spin Poincaré sphere. In isotropic Hermitian media for plasmonic or dielectric confinements of light, the emergence of a longitudinal field with an intrinsic $\pi/2$ phase difference with respect to the transverse field always accompanies the T-spin component. To annihilate the T-spin (Fig. 4b), the T-spin modes with different signs should be degenerate with the same amplitude ($\psi_- = \psi_+ \cdot e^{i\theta}$), where $\theta$ denotes the polarization angle with respect to the propagation direction. As shown in Supplementary Eq. (S3), the degeneracy of two T-spin modes $\psi_\pm$ with a phase difference $\theta$ is achieved with the complex anisotropic permittivity $\varepsilon_{x,y}$. Figures 4c,d show the required landscapes of complex potentials $\varepsilon_{x,y}$ for the zero T-spin modes with the profiles of Fig. 4e,f for positive polarization angles $\theta$. Almost all SOPs on the T-spin Poincaré sphere can be covered with spatially varying complex anisotropic media except for diverging states of $\theta = 0°$ (purely transversal wave) and $\theta = 180°$ (purely longitudinal wave). This T-spin Poincaré sphere in Figs 2-4 together with the classical Poincaré sphere completes the access to the '3-dimensional' polarization and spin eigenmodes with full degrees of freedom in a global regime. This 'annihilation' of the T-spin from complex potentials originating from the $\pi/2$ phase difference between transverse and longitudinal fields is distinct from the 'creation' of the longitudinal spin from parity-time-symmetric complex potentials[28].

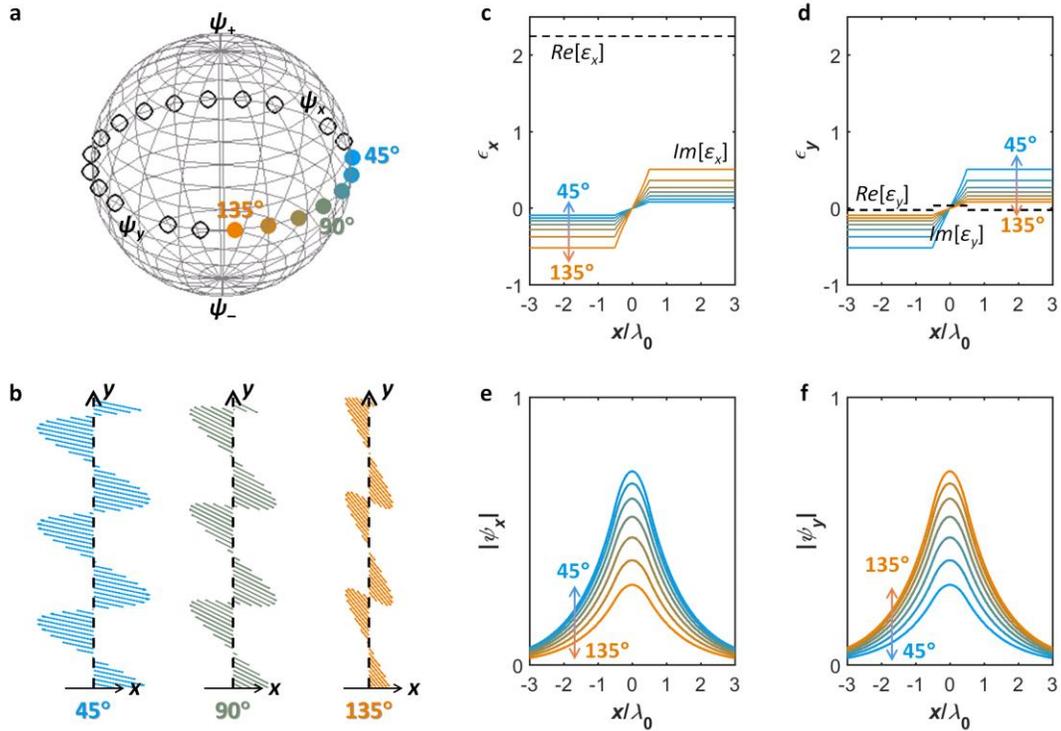

**Figure 4. Annihilation of T-spin modes in complex potentials.** (**a**) Representations of zero-T-spin modes on the T-spin Poincaré sphere (circles). The angles indicate the polarization angles determined by the phase difference between positive and negative T-spins with $\psi_-/\psi_+ = e^{i\theta}$: $\theta = 0°$ for the transverse wave and $\theta = 180°$ for the longitudinal wave. The colors of solid circles indicate the polarization angles used in (**c-f**). (**b**) Polarization evolution during the propagation is illustrated for $\theta = 45°$, 90°, and 135°. (**c,d**) Inversely designed complex anisotropic permittivities of (**c**) $\varepsilon_x$ and (**d**) $\varepsilon_y$ for the target eigenmodes with $\theta = 45°$ to 135°. Black dotted lines in (**c,d**) indicate the real part of the permittivities, and colored solid lines represent the imaginary part. (**e,f**) Magnitudes of the *x* and *y* polarizations (**e**) $|\psi_x|$ and (**f**) $|\psi_y|$ in the eigenmodes of $\theta = 45°$ to 135°. The target effective permittivity is $\varepsilon_e = 2.25$.

In conclusion, we demonstrated the global realization of T-spin eigenmodes by removing the inversion of the T-spin handedness in the spatial domain. To release the restriction, we apply a transition between hyperbolic and elliptic anisotropic materials around the uniaxial ENZ state. Based on the proposed platform, the density of the wavefunction itself becomes the observable of the T-spin mode, thus allowing far-field transport. With a novel polarization state of zero transverse spin, which represents in-phase oscillations between the transverse and longitudinal fields, all SOPs on the T-spin Poincaré sphere can be accessed. We expect that the full access to the T-spin Poincaré sphere demonstrated here will be used to stimulate generalized torque interactions[29,30] for photons.

**Methods Summary**

**Derivation of the T-spin interface mode** Consider the TM ($H_z$, $E_x$, $E_y$) interface mode between hyperbolic ($\varepsilon_x = \varepsilon_e + \Delta\varepsilon$ and $\varepsilon_y = -\Delta\varepsilon$ when $x < 0$) and elliptic ($\varepsilon_x = \varepsilon_e - \Delta\varepsilon$ and $\varepsilon_y = \Delta\varepsilon$ when $x > 0$) materials. For the functional form of the magnetic fields $H_z(x>0) = \psi_{h0} \cdot exp(-\alpha_1 \cdot x) \cdot exp(-i\beta y)$ and $H_z(x<0) = \psi_{h0} \cdot exp(\alpha_2 \cdot x) \cdot exp(-i\beta y)$ of the forward-propagating wave ($\beta > 0$), the boundary conditions for $H_z$ and $E_y$ yield a dispersion relation of

$$\beta = k_0 \cdot \sqrt{\varepsilon_e} \cdot \sqrt{1 - \left(\frac{\Delta\varepsilon}{\varepsilon_e}\right)^2}, \tag{2}$$

and $\alpha_1^2 = \alpha_2^2 = \alpha^2 = (\Delta\varepsilon^2/\varepsilon_e) \cdot k_0^2$. We can then obtain the ratio between the longitudinal ($E_y$) and transverse ($E_x$) electric fields,

$$\frac{E_y}{E_x} = \begin{array}{ll} i \cdot sgn(\Delta\varepsilon) \sqrt{\dfrac{\varepsilon_e - \Delta\varepsilon}{\varepsilon_e + \Delta\varepsilon}} & (x > 0) \\[2ex] i \cdot sgn(\Delta\varepsilon) \sqrt{\dfrac{\varepsilon_e + \Delta\varepsilon}{\varepsilon_e - \Delta\varepsilon}} & (x < 0) \end{array}, \tag{3}$$

which results in the T-spin mode $\Psi(x) = \psi_+(x) \cdot \mathbf{e}_+ + \psi_-(x) \cdot \mathbf{e}_-$, with

$$\psi_\pm(x) = \begin{array}{ll} \dfrac{\varepsilon_e + \Delta\varepsilon \pm sgn(\Delta\varepsilon) \cdot \sqrt{\varepsilon_e^2 - \Delta\varepsilon^2}}{2\varepsilon_e} \cdot \alpha \cdot e^{-\alpha x} & (x > 0) \\[2ex] \dfrac{\varepsilon_e - \Delta\varepsilon \pm sgn(\Delta\varepsilon) \cdot \sqrt{\varepsilon_e^2 - \Delta\varepsilon^2}}{2\varepsilon_e} \cdot \alpha \cdot e^{\alpha x} & (x < 0) \end{array}. \tag{4}$$

We note that $\Delta\varepsilon$ defines the trade-off between the strong confinement (large $\alpha$) and pure handedness ($||\psi_+|^2 - |\psi_-|^2| \gg 0$) of the T-spin mode. Although the discontinuity of $\Delta\varepsilon$ allows the confinement of the mode with $\alpha > 0$, it simultaneously breaks the pure T-spin condition from Eqs. (3) and (4). Moreover, the sign of $\Delta\varepsilon$, which is dependent on the spatial arrangement of hyperbolic and elliptic materials, determines the handedness of the mode via the term $sgn(\Delta\varepsilon)$.

**Acknowledgments**

This work was supported by the National Research Foundation of Korea (NRF) through the Global Frontier Program (GFP, 2014M3A6B3063708) funded by the Ministry of Science, ICT & Future Planning of the Korean government. X. Piao and N. Park were supported by the Korea Research Fellowship Program (KRF, 2016H1D3A1938069), and S. Yu was supported by the Basic Science Research Program (2016R1A6A3A04009723) through the NRF, and these programs are funded by the Ministry of Education of the Korean government.

**Author Contributions**

X.P. conceived the presented idea. X.P. and S.Y. developed the theory and performed the computations. N.P. encouraged X.P. to investigate the pure existence of transverse spins while supervising the findings of this work. All authors discussed the results and contributed to the final manuscript.

**Competing Interests Statement**

The authors declare that they have no competing financial interests.

# Supplementary Information for "Transverse spinning of light with globally unique handedness"

Xianji Piao, Sunkyu Yu, and Namkyoo Park*

*Photonic Systems Laboratory, Department of Electrical and Computer Engineering, Seoul National University, Seoul 08826, Korea*

*E-mail address for correspondence: nkpark@snu.ac.kr*

**Note S1**. Derivation of anisotropic optical potentials for T-spin modes

**Note S2**. Dependencies of T-spin modes on propagation directions and mirror symmetry

**Note S3.** Field distributions of Gaussian T-spin modes with different modal sizes

**Note S4.** Spatial profiles and polarizations of nearby eigenmodes

**Note S5.** Trade-off relation between the confinement and spin handedness

**Note S6.** Metamaterial realization of the T-spin interface

**Note S7.** Excitation conditions of the T-spin interface mode in metamaterials

**Note S1. Derivation of anisotropic optical potentials for T-spin modes**

Consider Maxwell's wave equation of the electric field vector $\mathbf{E} = E_x \cdot \mathbf{e_x} + E_y \cdot \mathbf{e_y}$, for 2-dimensional (2D) in-plane transverse magnetic (TM) modes in electrically anisotropic materials ($\varepsilon_{x,y}$):

$$\begin{bmatrix} -\partial_y^2 & \partial_y \partial_x \\ \partial_x \partial_y & -\partial_x^2 \end{bmatrix} \begin{bmatrix} E_x \\ E_y \end{bmatrix} = k_0^2 \cdot \begin{bmatrix} \varepsilon_x & 0 \\ 0 & \varepsilon_y \end{bmatrix} \begin{bmatrix} E_x \\ E_y \end{bmatrix}, \tag{S1}$$

where $k_0 = 2\pi/\lambda_0$ is the free space wavenumber. For 1-dimensional (1D) variations of optical potentials $\varepsilon_{x,y} = \varepsilon_{x,y}(x)$, the electric field can be set to $\mathbf{E}(x) = \mathbf{\Psi}(x) \cdot e^{-i\beta y}$ as $E_{x,y}(x,y) = \psi_{x,y}(x) \cdot e^{-i\beta y}$, where $\beta$ is the wavevector propagating toward the $y$-axis. By employing the T-spin representation $\mathbf{\Psi}(x) = \psi_+(x) \cdot \mathbf{e}_+ + \psi_-(x) \cdot \mathbf{e}_-$ to describe $z$-axis "transverse" spins to $y$-propagating waves as $\mathbf{e}_\pm = (\mathbf{e_x} \pm i \cdot \mathbf{e_y})/2^{1/2}$, Eq. (S1) then becomes

$$\begin{bmatrix} \beta^2 + \beta \cdot \partial_x & \beta^2 - \beta \cdot \partial_x \\ -i\beta \cdot \partial_x - i \cdot \partial_x^2 & -i\beta \cdot \partial_x + i \cdot \partial_x^2 \end{bmatrix} \begin{bmatrix} \psi_+ \\ \psi_- \end{bmatrix} = k_0^2 \cdot \begin{bmatrix} \varepsilon_x & \varepsilon_x \\ i\varepsilon_y & -i\varepsilon_y \end{bmatrix} \begin{bmatrix} \psi_+ \\ \psi_- \end{bmatrix}$$
$$= k_0^2 \cdot \begin{bmatrix} \psi_+ + \psi_- & 0 \\ 0 & i(\psi_+ - \psi_-) \end{bmatrix} \begin{bmatrix} \varepsilon_x \\ \varepsilon_y \end{bmatrix}. \tag{S2}$$

From Eq. (S2), the interaction between optical potentials and T-spin states of light can be represented as

$$\varepsilon_x = \varepsilon_e + \frac{\beta}{k_0^2} \cdot \frac{\partial_x(\psi_+ - \psi_-)}{\psi_+ + \psi_-}$$
$$\varepsilon_y = -\frac{1}{k_0^2} \cdot \frac{\partial_x^2(\psi_+ - \psi_-)}{\psi_+ - \psi_-} - \frac{\beta}{k_0^2} \cdot \frac{\partial_x(\psi_+ + \psi_-)}{\psi_+ - \psi_-}, \tag{S3}$$

where $\varepsilon_e$ is the eigenvalue (or effective permittivity) of the eigenmode $\mathbf{\Psi} = [\psi_+, \psi_-]^T$ from $\beta^2 = \varepsilon_e \cdot k_0^2$. Considering the $+y$-propagation only ($\beta > 0$), Eq. (S3) further reduces to

$$\varepsilon_x = \varepsilon_e + \frac{\sqrt{\varepsilon_e}}{k_0} \cdot \frac{\partial_x(\psi_+ - \psi_-)}{\psi_+ + \psi_-}$$
$$\varepsilon_y = -\frac{1}{k_0^2} \cdot \frac{\partial_x^2(\psi_+ - \psi_-)}{\psi_+ - \psi_-} - \frac{\sqrt{\varepsilon_e}}{k_0} \cdot \frac{\partial_x(\psi_+ + \psi_-)}{\psi_+ - \psi_-}. \tag{S4}$$

We note that contrary to the standard eigenvalue representation[1] of the wave equation deriving eigenstates from the given potential, Eq. (S3,S4) generates the required optical potential $\varepsilon_{x,y}(x)$ for the given set of the nodeless eigenmode $\mathbf{\Psi}$ and its eigenvalue $\varepsilon_e$.

Figure S1 shows the schematics of the detailed inverse design procedure of optical potentials defined by Eq. (S4). For the *y*-axis-propagating eigenmode of the arbitrary electric field profile (Fig. S1a,b), its envelope **Ψ** can be separated for each T-spin state as $\mathbf{\Psi}(x) = [\psi_+(x),\psi_-(x)]^T$ (Fig. S1c,d). The corresponding anisotropic optical potential $\varepsilon_{x,y}(x)$ is then obtained (Fig. S1e,f) from Eq. (S4) using **Ψ**(*x*) and the predefined eigenvalue $\varepsilon_e$. We note that the propagating mode with real-valued $\varepsilon_e$ leads to the dielectric-dominant field profile for the transverse permittivity $\varepsilon_x(x)$ (Fig. S1e) while the longitudinal permittivity $\varepsilon_y(x)$ has a rapidly varying distribution dependent on the spin states of light (Fig. S1f).

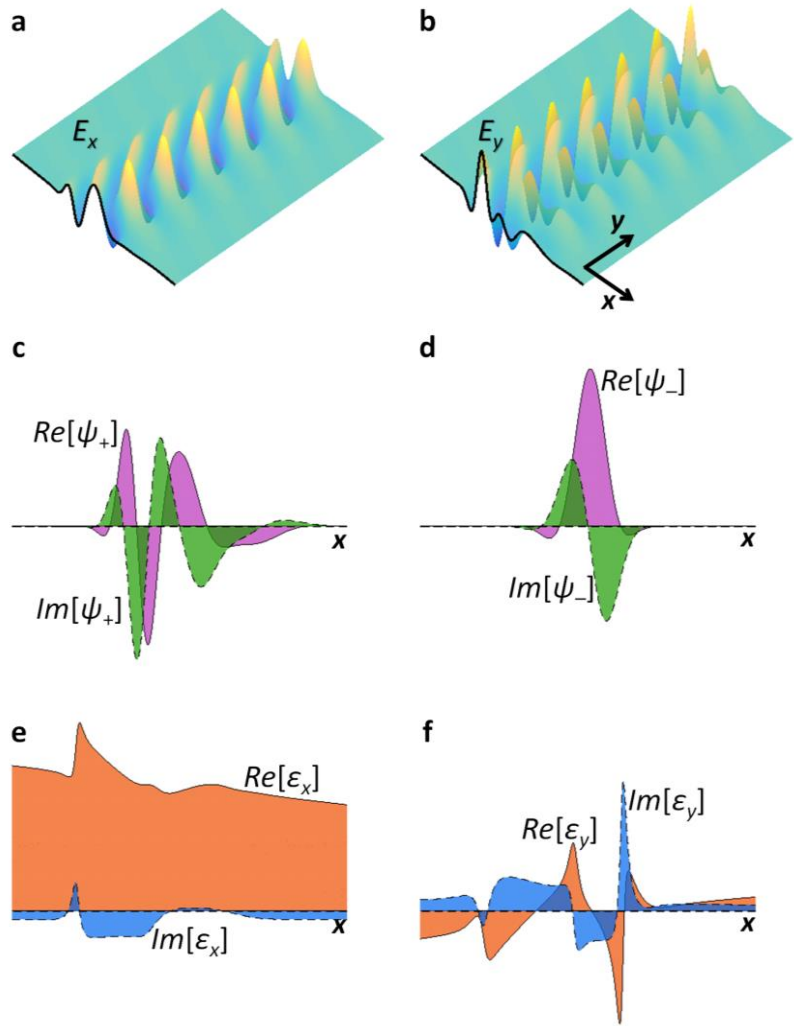

**Figure S1. Inverse design of optical transverse spins having arbitrary spatial distributions.** (**a**) Transverse and (**b**) longitudinal electric fields of the target eigenmode propagating along the *y*-axis. (**c,d**) Real (purple) and imaginary (green) components of (**c**) positive and (**d**) negative T-spin modes in (**a,b**). (**e,f**) Designed anisotropic permittivity (**e**) $\varepsilon_x$ and (**f**) $\varepsilon_y$ by Eq. (S4) for real (orange) and imaginary (blue) components.

**Note S2. Dependencies of T-spin modes on propagation directions and mirror symmetry**

The sign of $\beta$ in Eq. (S3) determines the landscape of T-spin potentials depending on the propagation direction ($\beta > 0$ for $+y$- and $\beta < 0$ for $-y$-propagation). For the globally pure T-spin modes of $\boldsymbol{\Psi}(x) = \psi_+(x) \cdot \mathbf{e}_+$ or $\boldsymbol{\Psi}(x) = \psi_-(x) \cdot \mathbf{e}_-$, the corresponding anisotropic potentials then become

$$\varepsilon_{x,(\pm y)}^{+} = \varepsilon_e \pm \frac{\sqrt{\varepsilon_e}}{k_0} \cdot \frac{\partial_x \psi_+}{\psi_+}$$

$$\varepsilon_{y,(\pm y)}^{+} = -\frac{1}{k_0^2} \cdot \frac{\partial_x^2 \psi_+}{\psi_+} \mp \frac{\sqrt{\varepsilon_e}}{k_0} \cdot \frac{\partial_x \psi_+}{\psi_+}$$

(S5)

for the forward ($+y$) and backward ($-y$) modes of $\boldsymbol{\Psi}(x) = \psi_+(x) \cdot \mathbf{e}_+$ (+ superscript in Eq. (S5)), and

$$\varepsilon_{x,(\pm y)}^{-} = \varepsilon_e \mp \frac{\sqrt{\varepsilon_e}}{k_0} \cdot \frac{\partial_x \psi_-}{\psi_-}$$

$$\varepsilon_{y,(\pm y)}^{-} = -\frac{1}{k_0^2} \cdot \frac{\partial_x^2 \psi_-}{\psi_-} \pm \frac{\sqrt{\varepsilon_e}}{k_0} \cdot \frac{\partial_x \psi_-}{\psi_-}$$

(S6)

for the forward ($+y$) and backward ($-y$) modes of $\boldsymbol{\Psi}(x) = \psi_-(x) \cdot \mathbf{e}_-$ (− superscript in Eq. (S6)).

Equations (S5) and (S6) represent the relations of (i) $\varepsilon_{x,(+y)}^{+} = \varepsilon_{x,(-y)}^{-}$, $\varepsilon_{y,(+y)}^{+} = \varepsilon_{y,(-y)}^{-}$ and (ii) $\varepsilon_{x,(+y)}^{-} = \varepsilon_{x,(-y)}^{+}$, $\varepsilon_{y,(+y)}^{-} = \varepsilon_{y,(-y)}^{+}$ if the target spatial profiles of T-spin modes are identical as $\psi_+(x) = \psi_-(x)$. These relations demonstrate that 'any' T-spin potential designed by Eq. (S5) or (S6) has the forward and backward eigenmodes of the 'opposite' T-spin but with the identical spatial profile (Fig. S2a,b). Moreover, when $\psi_\pm(x)$ are even functions (or $\partial_x\psi_\pm(x)$ are odd functions), the potentials for each T-spin satisfy the relations $\varepsilon_{x,(+y)}^{+}(x) = \varepsilon_{x,(+y)}^{-}(-x)$ and $\varepsilon_{y,(+y)}^{+}(x) = \varepsilon_{y,(+y)}^{-}(-x)$, which implies that the mirror-symmetric potential possesses the 'opposite' T-spin eigenmode.

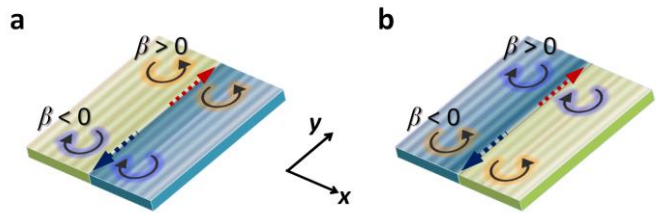

**Figure S2. Dependencies of T-spin potentials on the propagation direction and mirror symmetry of the target eigenmode $\psi_\pm(x)$.** (**a,b**) T-spin modes of forward and backward waves obtained from Eqs. (S5) and (S6). The target eigenmodes are assumed to have an even parity. Solid arrows indicate T-spin modes. Dotted arrows depict the propagation direction.

## Note S3. Field distributions of Gaussian T-spin modes with different modal sizes

Figure S3 shows the detailed field profiles ($H_z$, $E_x$, $E_y$) of the Gaussian T-spin wave (Fig. S3a,b: $\sigma = 5\lambda_0$ and Fig. S3c,d: $\sigma = \lambda_0/5$). The field amplitudes of $E_x$ and $E_y$ have identical Gaussian distributions with a $\pi/2$ phase difference: $E_x(x,y) = -iE_y(x,y) = \exp(-x^2/(2\sigma^2)) \cdot e^{-i\beta y}$. The magnetic field then becomes $H_z(x,y) = -[(k_0 \cdot \varepsilon_e^{1/2} - x/\sigma^2)/(\omega\mu_0)] \cdot E_x(x,y)$, which is composed of symmetric ($-k_0 \cdot \varepsilon_e^{1/2} \cdot E_x/(\omega\mu_0)$) and antisymmetric ($(x/\sigma^2) \cdot E_x(x,y)/(\omega\mu_0)$) parts.

For wavelength-scale confinement (Fig. S3a,b for $\sigma = 5\lambda_0$), the contribution of the symmetric part is dominant ($k_0 \cdot \varepsilon_e^{1/2} \gg 1/\sigma$), which leads to nearly symmetric $H_z$ (Fig. S3a) and Poynting vector $P_y$ (Fig. S3b) profiles. In contrast, subwavelength confinement ($\sigma \ll \lambda_0$) increases the contribution from the antisymmetric part of $H_z$ (Fig. S3c,d for $\sigma = \lambda_0/5$, $k_0 \cdot \varepsilon_e^{1/2} \sim 1/\sigma$), which results in asymmetric $H_z$ (Fig. S3c) and $P_y$ (Fig. S3d) profiles.

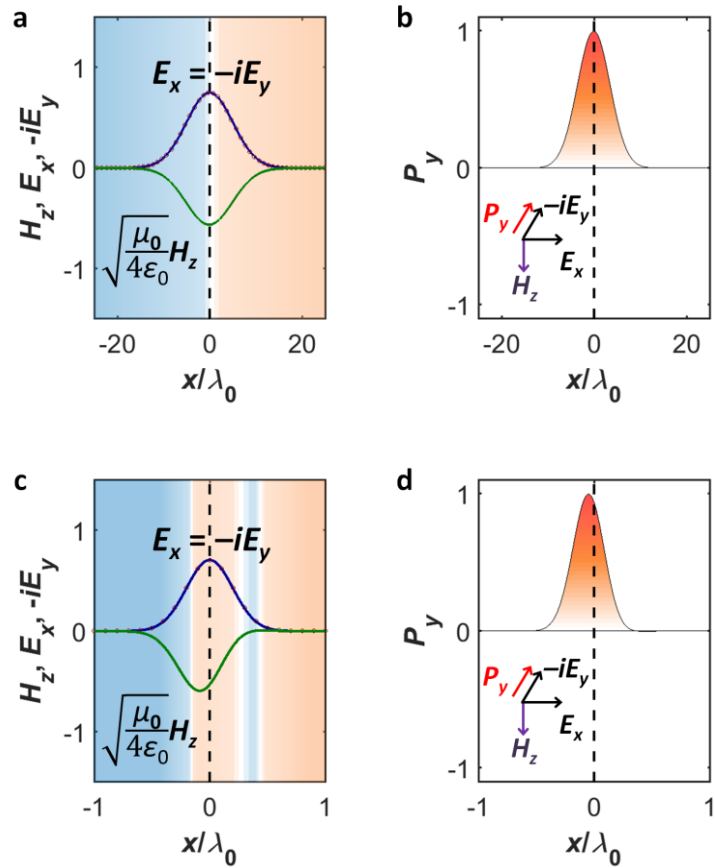

**Figure S3. Detailed field profiles and Poynting vectors of Gaussian T-spin modes.** (**a,c**) The TM mode fields ($H_z$, $E_x$, $E_y$) and (**b,d**) Poynting vectors ($P_y$) for (**a,b**) $\sigma = 5\lambda_0$ and (**c,d**) $\sigma = \lambda_0/5$. Background colors in (**a,c**) indicate the topology of materials (blue for indefinite media and orange for definite media). All other parameters are the same as those in Fig. 2 in the main text.

**Note S4. Spatial profiles and polarizations of nearby eigenmodes**

Figure S4 presents the eigenmodes (Fig. S4a,b,e,f) near the target T-spin eigenmode (Fig. S4c,d). The lower-index mode ($\varepsilon_e = 2.16$) is concentrated in the definite material (Fig. S4a), whereas the higher-index mode ($\varepsilon_e = 2.39$) is concentrated in the indefinite material (Fig. S4e). Such a spatial separation for different eigenmodes can be understood through the spatial distribution of the IFC (Fig. 2c in the main text), which shows a low $k$ for the definite region and high $k$ for the indefinite region. The polarization of nearby eigenmodes is close to linear (Fig. S4b,f), demonstrating the selective excitation of the target T-spin eigenmode (Fig. S4d).

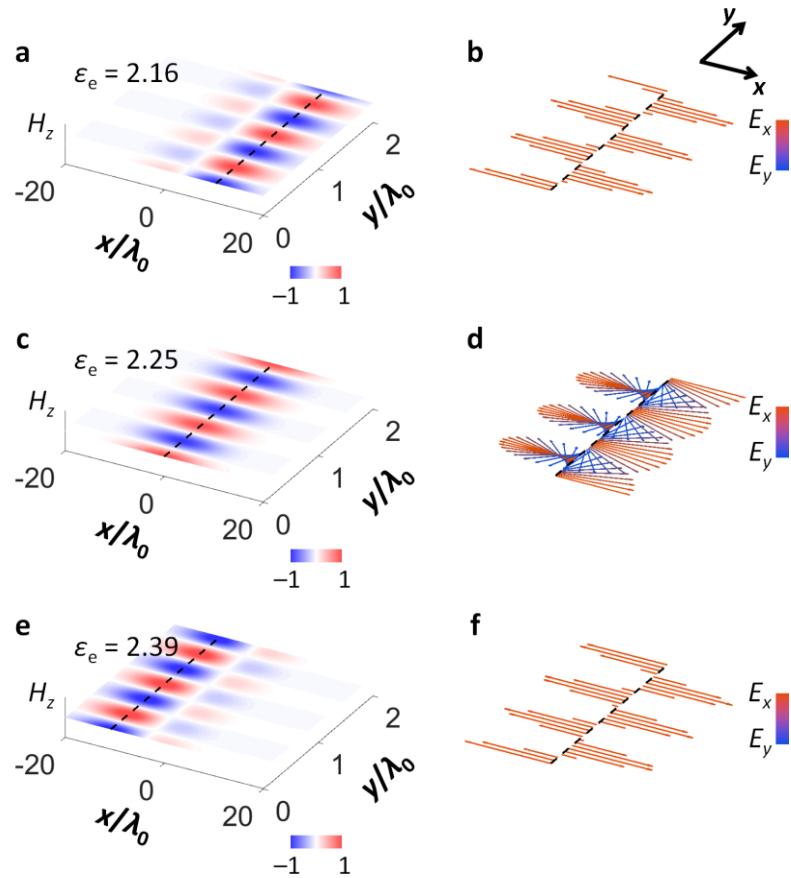

**Figure S4. Spatial separation of eigenmodes in the designed T-spin optical potential.** Three eigenmodes around the target effective permittivity $\varepsilon_e = 2.25$ for (**a,b**) the lower-index mode of $\varepsilon_e = 2.16$, (**c,d**) the target T-spin mode of $\varepsilon_e = 2.25$, and (**e,f**) the higher-index mode of $\varepsilon_e = 2.39$. The color bars in (**b,d,f**) present the ratio of transverse ($E_x$) and longitudinal ($E_y$) electric fields. All other parameters are the same as those in Fig. 2 in the main text.

## Note S5. Trade-off relation between the confinement and spin handedness

The modal size of evanescent waves can be evaluated from[2] $A = \int W(x)dx / max\{W(x)\}$, where $W(x) = (\mathbf{D}\cdot\mathbf{E} + \mathbf{B}\cdot\mathbf{H})/2$ is the electromagnetic energy of the mode. Figure S5a shows the variation of the modal size of T-spin interface modes with respect to $\Delta\varepsilon$. The increased discontinuity by $|\Delta\varepsilon|$ exponentially reduces the modal size, which is expected from the dispersion relation of the decay constant $\alpha^2 = (\Delta\varepsilon^2/\varepsilon_e)\cdot k_0^2$ in the Methods Summary.

As shown in Eq. (3) in the Methods Summary, $\Delta\varepsilon$ breaks the pure spin condition $E_y = \pm iE_x$. We quantify the T-spin purity of the T-spin interface mode by $S = \int(|\psi_+|^2 - |\psi_-|^2)dx / \int(|\psi_+|^2 + |\psi_-|^2)dx$, where $S = +1$ for purely positive spin and $S = -1$ for purely negative spin. Although the increase of the discontinuity by $|\Delta\varepsilon|$ decreases the spin purity, nearly ideal T-spin waves can be achieved except in the case of deep-subwavelength confinement (Fig. S5b, $S \sim \pm 0.993$ for $A \sim \lambda_0$ and $S \sim \pm 0.829$ for $A \sim \lambda_0/5$).

The confinement of the T-spin density can also be defined as $S\lambda_0/A$ (Fig. S5c), which represents the 'concentration' of the spin state for each wavelength. For a given value of $\varepsilon_e$, an optimum point occurs for T-spin confinement ($\Delta\varepsilon = 1.42$ in Fig. S5c).

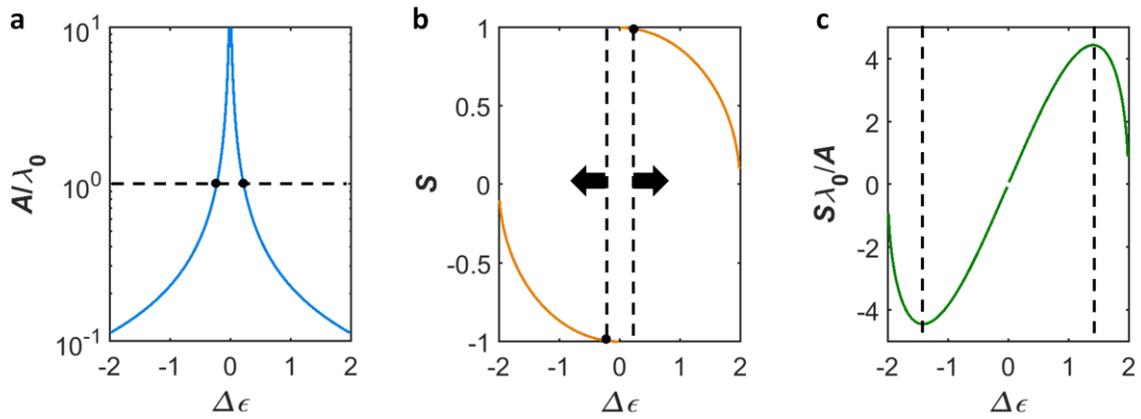

**Figure S5. Trade-off relation between the modal size and spin purity of the T-spin interface mode:** (a) modal size $A/\lambda_0$, (b) spin purity $S$, and (c) spin confinement $S\lambda_0/A$. Black dashed lines in (a,b) indicate the wavelength-size confinement, and black arrows in (b) show the regime of subwavelength confinement. Black dashed lines in (c) present the optimum confinement of the T-spin. All other parameters are the same as those in Fig. 3 in the main text.

## Note S6. Metamaterial realization of the T-spin interface

To implement the interface between hyperbolic ($\varepsilon_x = \varepsilon_e + \Delta\varepsilon$ and $\varepsilon_y = -\Delta\varepsilon$) and elliptic ($\varepsilon_x = \varepsilon_e - \Delta\varepsilon$ and $\varepsilon_y = \Delta\varepsilon$) materials, we apply the effective medium theory[3,4] of subwavelength multilayered structures (Fig. 3c in the main text). For the unit cell composed of alternating layers of $\varepsilon_{1H}$ and $\varepsilon_{2H}$ (or $\varepsilon_{1E}$ and $\varepsilon_{2E}$) in a hyperbolic (or an elliptic) region, the anisotropic effective permittivities become

$$\varepsilon_{x,H} = \frac{1}{\frac{f_H}{\varepsilon_{1H}} + \frac{1-f_H}{\varepsilon_{2H}}} = \varepsilon_e + \Delta\varepsilon,$$

$$\varepsilon_{y,H} = \varepsilon_{1H} \cdot f_H + \varepsilon_{2H} \cdot (1-f_H) = -\Delta\varepsilon \quad (S7)$$

for the hyperbolic region, and

$$\varepsilon_{x,E} = \frac{1}{\frac{f_E}{\varepsilon_{1E}} + \frac{1-f_E}{\varepsilon_{2E}}} = \varepsilon_e - \Delta\varepsilon,$$

$$\varepsilon_{y,E} = \varepsilon_{1E} \cdot f_E + \varepsilon_{2E} \cdot (1-f_E) = \Delta\varepsilon \quad (S8)$$

for the elliptic region, where $f_H$ and $f_E$ denote the filling ratios of the media $\varepsilon_{1H}$ and $\varepsilon_{1E}$, respectively.

Figure S6 represents the required permittivities of each layer in hyperbolic (Fig. S6a) and elliptic (Fig. S6b) materials for $\Delta\varepsilon = 0.5$. We adopt the case of $f_H = 0.7$ ($\varepsilon_{1H} = 1.5$ and $\varepsilon_{2H} = -5.2$) and $f_E = 0.9$ ($\varepsilon_{1E} = 1.3$ and $\varepsilon_{2E} = -6.9$) for the results of Fig. 3 in the main text, which allows for the utilization of dielectric and metallic layers with experimentally accessible material parameters.

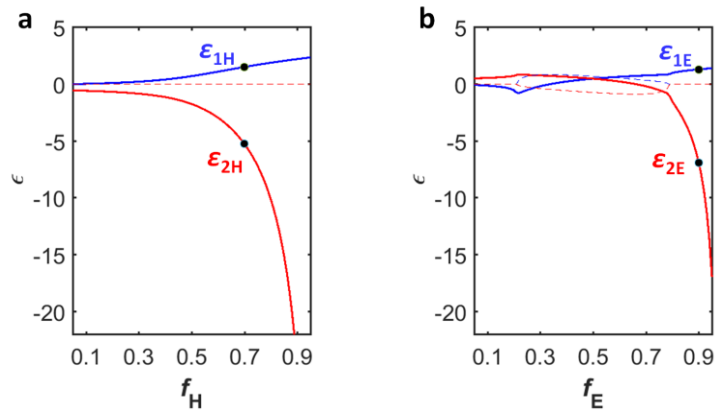

**Figure S6. Required permittivities of each layer for $\Delta\varepsilon = 0.5$ as a function of the filling ratio of $\varepsilon_{1H}$ and $\varepsilon_{1E}$:** in (**a**) hyperbolic and (**b**) elliptic materials. Solid and dashed lines indicate the real and imaginary parts of the permittivity, respectively. All other parameters are the same as those in Fig. 3 in the main text. Solid dots represent the parameter values used for Fig. 3 in the main text.

**Note S7. Excitation conditions of the T-spin interface mode in metamaterials**

For an oblique incidence (Fig. 3c in the main text), the incident light requires the transverse wavevector to be same as the propagation constant of the T-spin interface mode $\beta$. From Eq. (2) in the Methods Summary, the excitation angle then becomes

$$\varphi = \arcsin\left(\sqrt{\frac{\varepsilon_e}{\varepsilon_b}\cdot\left[1-\left(\frac{\Delta\varepsilon}{\varepsilon_e}\right)^2\right]}\right). \tag{S9}$$

where $\varepsilon_b$ is the permittivity of the incident region. Without introducing structural modulations[5] or near-field techniques[6], the permittivity of the incident region is restricted by $\varepsilon_b > \varepsilon_e\cdot[1 - (\Delta\varepsilon/\varepsilon_e)^2]$, which is similar to the prism coupling technique for the excitation of surface plasmon polaritons[7].

Figure S7 presents the S-matrix-calculated[8] angular dependency of the excited T-spin mode energy at the interface for unit cell lengths of $\lambda_0/5$, $\lambda_0/10$, $\lambda_0/20$, and $\lambda_0/40$. As shown, the deep-subwavelength layer construction of metamaterial leads to an optimum angle close to that of the theoretical value. The trade-off relation between the excited energy and angular bandwidth is also observed.

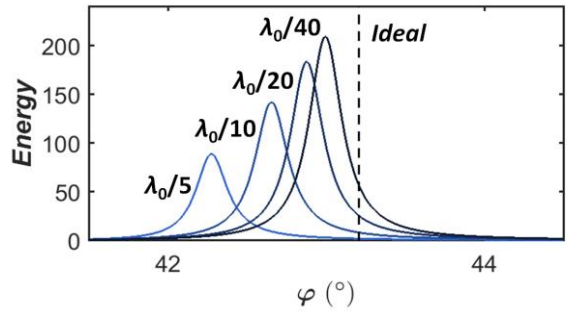

**Figure S7. Angular dependency of the excited energy at the interface for different conditions of metamaterial realization:** unit cell lengths of $\lambda_0/5$, $\lambda_0/10$, $\lambda_0/20$, and $\lambda_0/40$. The results are calculated by the S-matrix technique[8], which is stable for the analysis of evanescent waves. The energy is normalized by the intensity of the incident wave. The dashed line indicates the excitation angle in the ideal case. All other parameters are the same as those in Fig. 3 in the main text.